\documentclass[sigconf]{acmart}

\usepackage{tabularx, booktabs}
\usepackage{multirow}
\usepackage{float} 
\usepackage{balance}
\usepackage{amsmath}
\usepackage[ruled,linesnumbered]{algorithm2e}
\usepackage{graphicx}
\usepackage{subfigure}
\usepackage{epstopdf}
\usepackage{tablefootnote}
\usepackage[flushleft]{threeparttable}
\usepackage{diagbox}
\usepackage{booktabs}
\usepackage{siunitx}
\usepackage{footmisc}
\usepackage{footnote}
\usepackage{enumitem}
\usepackage[flushleft]{threeparttable}
\usepackage{etoolbox}
\appto\TPTnoteSettings{\footnotesize}

\usepackage{tablefootnote}

\copyrightyear{2023}
\acmYear{2023}
\setcopyright{acmlicensed}
\acmConference[SIGIR '23] {Proceedings of the 46th International ACM SIGIR Conference on Research and Development in Information Retrieval}{July 23--27, 2023}{Taipei, Taiwan.}
\acmBooktitle{Proceedings of the 46th International ACM SIGIR Conference on Research and Development in Information Retrieval (SIGIR '23), July 23--27, 2023, Taipei, Taiwan}
\acmISBN{978-1-4503-9408-6/23/07}
\begin{document}

\title{Constructing Tree-based Index for Efficient and Effective Dense Retrieval}

\author{Haitao Li}
\affiliation{DCST, Tsinghua University}
\affiliation{Zhongguancun Laboratory}
\email{liht22@mails.tsinghua.edu.cn}

\author{Qingyao Ai}
\affiliation{DCST, Tsinghua University}
\affiliation{Zhongguancun Laboratory}
\email{aiqy@tsinghua.edu.cn}

\author{Jingtao Zhan}
\affiliation{DCST, Tsinghua University}
\affiliation{Zhongguancun Laboratory}
\email{chenjia0831@gmail.com}

\author{Jiaxin Mao}
\affiliation{GSAI, Renmin University of China}
\email{maojiaxin@gmail.com}

\author{Yiqun Liu}
\authornote{Corresponding author}
\affiliation{DCST, Tsinghua University}
\affiliation{Zhongguancun Laboratory}
\email{yiqunliu@tsinghua.edu.cn}

\author{Zheng Liu}
\affiliation{Huawei Poisson Lab}
\email{liuzheng107@huawei.com}

\author{Zhao Cao}
\affiliation{Huawei Poisson Lab}
\email{caozhao1@huawei.com}

\renewcommand{\shortauthors}{Li, et al.}

\begin{abstract}
Recent studies have shown that Dense Retrieval (DR) techniques can significantly improve the performance of first-stage retrieval in IR systems. Despite its empirical effectiveness, the application of DR is still limited. In contrast to statistic retrieval models that rely on highly efficient inverted index solutions, DR models build dense embeddings that are difficult to be pre-processed with most existing search indexing systems. To avoid the expensive cost of brute-force search, the Approximate Nearest Neighbor (ANN) algorithm and corresponding indexes are widely applied to speed up the inference process of DR models. Unfortunately, while ANN can improve the efficiency of DR models, it usually comes with a significant price on retrieval performance. 

To solve this issue, we propose JTR, which stands for \textbf{J}oint optimization of \textbf{TR}ee-based index and query encoding. Specifically, we design a new unified contrastive learning loss to train tree-based index and query encoder in an end-to-end manner. The tree-based negative sampling strategy is applied to make the tree have the maximum heap property, which supports the effectiveness of beam search well. Moreover, we treat the cluster assignment as an optimization problem to update the tree-based index that allows overlapped clustering. We evaluate JTR on numerous popular retrieval benchmarks. Experimental results show that JTR achieves better retrieval performance while retaining high system efficiency compared with widely-adopted baselines. It provides a potential solution to balance efficiency and effectiveness in neural retrieval system designs. 
\end{abstract}



\maketitle

\section{Introduction}

Information Retrieval (IR) system has become one of the most important tools for people to find useful information online. In practice, an industry IR system usually needs to retrieve a small set of relevant documents from millions or even billions of documents. The computational cost of brute-force search is usually unacceptable due to the system latency requirements. Therefore, indexes are often employed to achieve a fast response. One of the most well-known examples is the inverted index~\cite{IVF}. Unfortunately, such traditional indexing techniques cannot be applied to recently proposed Dense Retrieval (DR) models. As DR models have shown promising performance in multi-stage retrieval systems, especially in the first stage retrieval~\cite{lin2020distilling,qu2020rocketqa,zhan2020repbert,karpukhin2020dense,khattab2020colbert,xie2023t2ranking}, how to develop effective and efficient indexing technique for dense retrieval has become an important question for the research community.

Efficient DR solutions such as Approximate Nearest Neighbor (ANN) search algorithms have been widely applied in the first stage retrieval process. Popular methods include tree-based indexes ~\cite{Annoy}, locality sensitive hashing (LSH)~\cite{johnson2019billion}, product quantization (PQ)~\cite{jegou2010product,OPQ}, hierarchical navigable small-world network (HNSW)~\cite{HNSW}, etc. Since tree-based indexes can achieve sub-linear time complexity by pruning low-quality candidates, it has drawn significant attention.

Despite success in improving retrieval efficiency, most existing tree-based indexes often come with the degradation of retrieval performance. There are two main reasons: (1) The majority of existing indexes cannot benefit from supervised data because they use task-independent reconstruction error as the loss function. (2) The training objectives of the index structure and the encoder are inconsistent. In general, the optimization goal of index structure is to minimize the approximation error, while the optimization goal of the encoder is to get better retrieval performance. This inconsistency may lead to sub-optimal results.

To balance the effectiveness and efficiency of the tree-based indexes, we propose JTR, which stands for \textbf{J}oint optimization of \textbf{TR}ee-based index and query encoding. To jointly optimize index structure and query encoder in an end-to-end manner, JTR drops the original ``encoding-indexing" training paradigm and designs a unified contrastive learning loss. However, training tree-based indexes using contrastive learning loss is non-trivial due to the problem of differentiability. To overcome this obstacle, the tree-based index is divided into two parts: cluster node embeddings and cluster assignment. For differentiable cluster node embeddings, which are small but very critical, we design tree-based negative sampling to optimize them. For cluster assignment, an overlapped cluster method is applied to iteratively optimize it.

To verify the effectiveness of our method, we compared JTR with a wide range of ANN methods on two large publicly available datasets. The empirical experimental results show that JTR achieves better performance compared to baselines while keeping high efficiency. Through the ablation study, we have confirmed the effectiveness of the proposed strategies\footnote{Code are available at \url{https://github.com/CSHaitao/JTR}.}.

In short, our main contributions are as follows:
\begin{enumerate} 
\item We propose a novel tree-based index for Dense Retrieval. Benefiting from the tree structure, it achieves sub-linear time complexity while still yielding promising results.  
\item We propose a new joint optimization framework to learn both the tree-based index and query encoder. To the best of our knowledge, JTR is the first joint-optimized retrieval approach with tree-based index. This framework improves end-to-end retrieval performance through unified contrastive learning loss and tree-based negative sampling.
\item We relax the constraint that documents are mutually exclusive between clusters and propose an efficient clustering solution in which a document can be assigned into multiple clusters to optimize the cluster assignment. We demonstrate that overlapped cluster is beneficial for improving retrieval performance.
\end{enumerate}


\section{Related Work}
\subsection{Dual Encoders}
With the development of Pre-trained Language Models (PLMs), Dense Retrieval (DR) has developed rapidly in the last decade. DR typically uses dual encoders to obtain embeddings of queries and documents and utilizes inner products as similarity factors.
For a query-document pair $<q,d>$, the primary goal is to learn a presentation function $f(\cdot)$ that maps text to higher-dimensional embeddings. The general form of a dual encoder is as follows:
\begin{equation}\label{eqn-1} 
  s(q,d)=<f(q),f(d)>
\end{equation}
Where, $<,>$ refers to the inner product. $s(q,d)$ is the relevance score. To further improve the performance of dual encoders, many researchers have improved the loss function, sampling method, and so on to make them more suitable for information retrieval tasks~\cite{gao2021complement, STAR,chenthuir,yang2022thuir,dong2022incorporating,yang2023enhance,li2023towards,lee2019contextualized}. For example, Lee et al.~\cite{lee2019contextualized} suggest the use of random negative sampling in large batches, which increases the difficulty of training. Zhan et al.~\cite{STAR} propose the dynamic hard negative sampling method, which effectively improves the performance of the DR model.

\subsection{Approximate Nearest Neighbor Search}
When queries and documents are encoded into embeddings, the retrieval problem can be considered as a Nearest Neighbor (NN) search problem. The simplest nearest neighbor search method is brute-force search, which becomes impractical when the corpus size explodes. Therefore, most studies use Approximate Nearest Neighbor (ANN) search.
In general, there are four common methods, including tree-based methods~\cite{Annoy}, hashing~\cite{johnson2019billion,FALCONN}, quantization-based methods~\cite{jegou2010product,OPQ}, and graph-based methods~\cite{HNSW}. Tree-based and graph-based methods divide embeddings into different spaces and query embeddings are retrieved only in similar spaces. Hashing and quantization-based methods compress vector representations in different ways. Both are important and are often used in combination in practice.

\subsection{Joint Optimization}
Existing DR systems follow the ``encoding-indexing" paradigm, Inconsistent optimization objectives of the two steps during training will lead to sub-optimal performance. Previous studies have shown that the combination of optimized index construction and retrieval model training can make the index directly benefit from the annotation information and achieve better retrieval performance~\cite{JPQ,RepCONC,TDM,feng2022forest,feng2022recommender,fang2022joint}. Zhan et al~\cite{JPQ} explore joint optimization of query encoding and product quantization, achieving state-of-the-art results. Furthermore, RepCONC~\cite{RepCONC} treats quantization as a constrained clustering process, requiring vectors to be clustered around quantized centroids. These approaches achieve good vector compression results, but the retrieval time complexity is still linear with respect to the corpus size. Tree-based methods have shown promising performance in recommender systems~\cite{TDM,TDM2,TDM3}. For example, TDM~\cite{TDM} proposes a tree-based deep recommendation model, which can be incorporated into any high-level model for retrieval. However, these methods are designed for recommendation systems and cannot be directly applied to retrieval systems.

Recently, Tay et al.~\cite{DSI} regarded transformer memory as the Differentiable Search Index (DSI). DSI maps documents into document IDs when indexing, and generates the corresponding docid for query when retrieving, so as to unify the training and indexing process. On this basis, Wang et al.~\cite{NCI} propose the Neural Corpus Indexer (NCI), which supports end-to-end document retrieval through a sequence-to-sequence network. These mode-based indexes provide a new paradigm for learning end-to-end retrieval systems. However, they are not suitable for larger-scale web searches. And when adding or deleting documents to the system, it is difficult to update the index. Compared with them, this paper focuses on the joint optimization of index and encoders, which preserves the advantages of the original index structure.


\begin{table}[t]
        \vspace{-0.15in}
	\caption{Important notations present in this paper.}
        \vspace{-0.15in}
        \label{notation}
	\begin{tabular}{p{0.1\columnwidth}p{0.85\columnwidth}}
		\toprule	
		\underline{\textbf{\emph{Common Notation}}} \\
		$q$ & a specific query  \\ 
		$d$ & a specific document \\
            $n$ & a specific cluster node \\
            $D$ & the document corpus, $D=\{ d_1, d_2,...,d_n \}$\\
            $\mathbb{D}$ & dimension of dense embeddings \\
            $\widetilde{e}_{c}$ & embedding of cluster nodes \\
            $\widetilde{e}_{d}$ & embedding of documents \\
            $S^{i}$ & documents set assigned to leaf node $i$ \\
            $\Phi(\cdot)$ & the query encoder \\
            \underline{\textbf{\emph{Tree Parameters}}}  \\
            $\beta$ & branch balance factor, representing the number of child nodes of non-leaf nodes  \\
            $\gamma$ & leaf balance factor, representing the maximum number of documents contained by one leaf node \\
            $b$ & beam size, indicating that the top $b$ leaf nodes will be retrieved \\
            \underline{\textbf{\emph{Overlapped Cluster Parameters}}}  \\

            $\mathbf{L}$ & the number of quries in the training set  \\
            $\mathbf{K}$ & the number of leaf nodes \\
            $\mathbf{N}$ & the number of documents \\
            $\mathbb{M}$ & matrix $\mathbb{M}\in \{0,1\}^{\mathbf{L} \times \mathbf{K}}$, which represents the relationship between the training queries and the leaf nodes \\ 
            $\mathbb{Y}$ & matrix $\mathbb{Y}\in \{0,1\}^{\mathbf{L} \times \mathbf{N}}$, which indicates the relevance of the queries and documents\\
            $\mathbb{C}$ & matrix $\mathbb{C}\in \{0,1\}^{\mathbf{N} \times \mathbf{K}}$, which represents the relationship between documents and leaf nodes\\
            $\lambda$ & number of repeated documents, which means that one document appears at most $\lambda$ times in the leaf nodes after overlapped cluster \\    
			
        \bottomrule
	\end{tabular}
\vspace{-3mm}
\end{table}

\section{THE JTR MODEL}
In this section, we first introduce the preliminary and tree structure in JTR. Secondly, the end-to-end joint optimization process is described in detail. Finally, we show how to optimize the cluster assignments of documents. Table~\ref{notation} shows the important notations that are present in this paper.

\begin{figure}[ht]
\vspace{-0.15in}
\centerline{\includegraphics[width=\columnwidth]{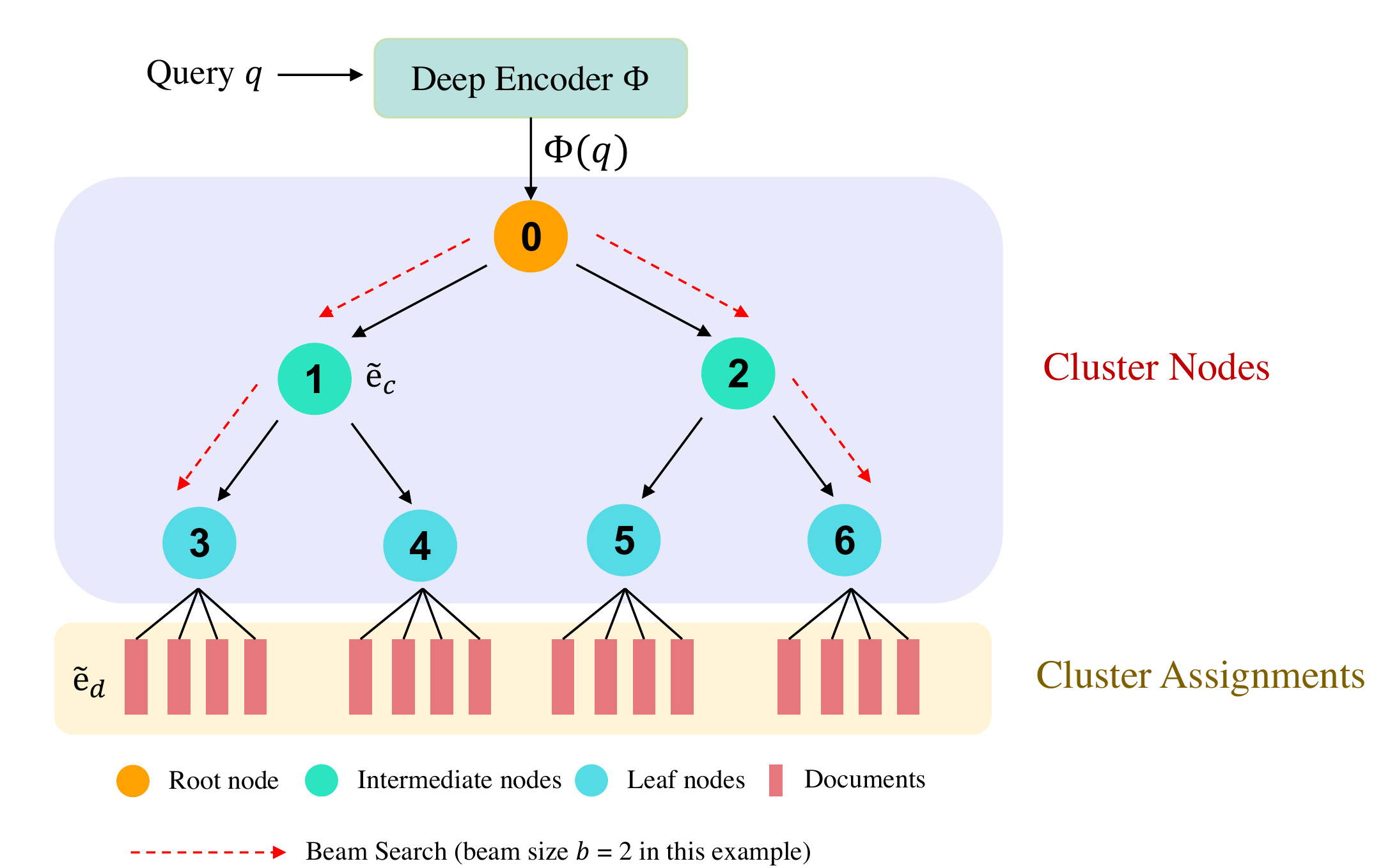}}
\vspace{-0.15in}
\caption{Illustration of the JTR tree structures. The integer represents the sequence number of the node. In this case, The tree has a depth of 3, number of clusters 4, branch balance factor $\beta$ = 2, and leaf balance factor $\gamma$ = 4. The beam size $b$ is set to 2.}
\vspace{-4mm}
\label{fig1}
\end{figure}

\begin{figure}[t]
\centerline{\includegraphics[width=0.8\columnwidth]{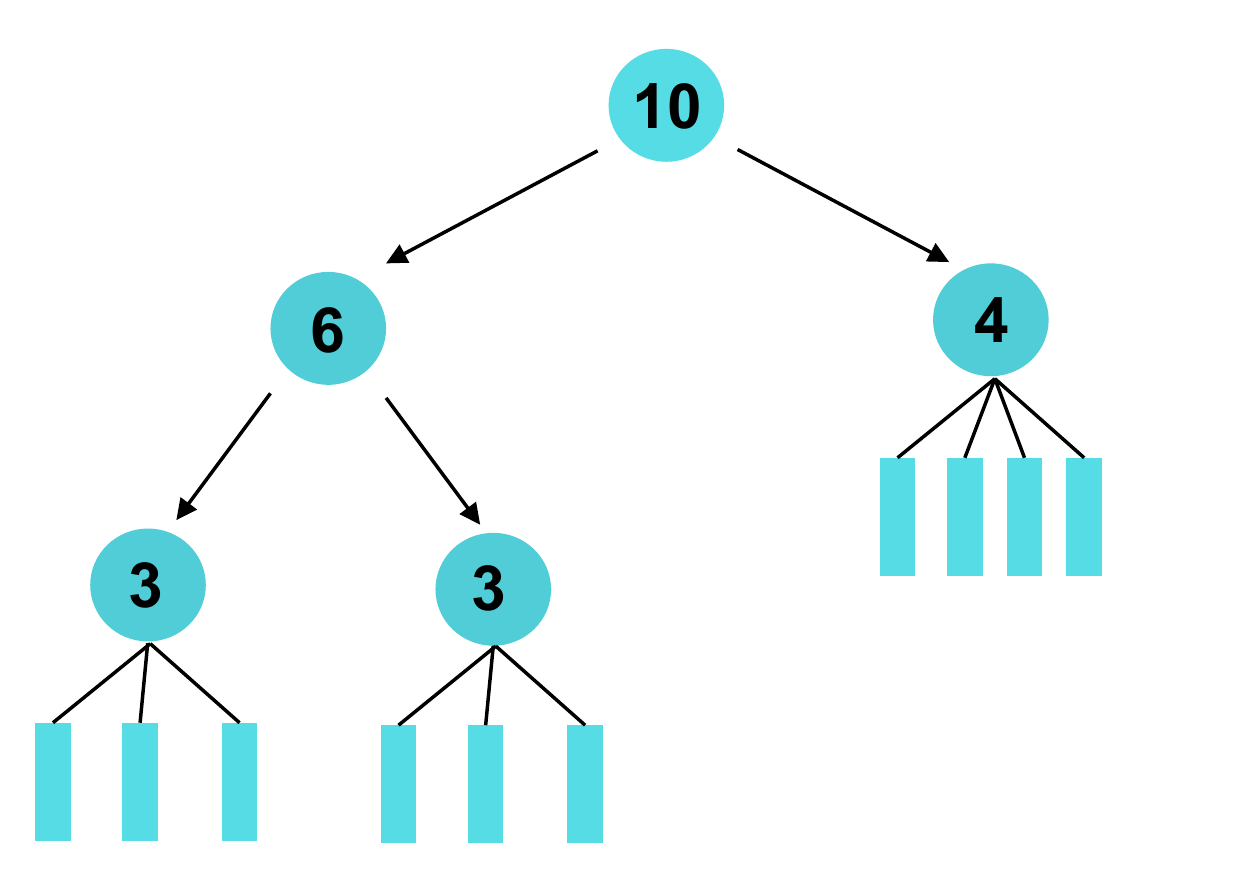}}
\vspace{-0.15in}
\caption{Initialization of the tree structure. The integer in nodes indicates the number of documents the node contains. In this case, the total number of documents is 10, the branch balance factor $\beta$ = 2, and the leaf balance factor $\gamma$ = 5. If the node contains more documents than $\gamma$, then k-means will be performed on the document embedding in the node until all nodes contain less than $\gamma$ documents. The embedding of each node is initialized as the cluster centroid embedding.}
\vspace{-5mm}
\label{cluster}
\end{figure}

\subsection{Preliminary}
Dense Retrieval can be formalized as follows: given a query $q$ and the corpus $D=\{ d_1, d_2,...,d_n \}$, the model needs to retrieve the top $k$ most relevant documents from $D$ with $q$. The training set is given in the form of ${\{(q_i,d_i)...\}}$, which means that $q_i$ and $d_i$ are relevant.

Tree-based index clusters all documents and prunes irrelevant cluster nodes at retrieval time to improve retrieval efficiency. As depicted in Figure \ref{fig1}, components in JTR are as follows:
\begin{itemize}[leftmargin=*]

    \item \textbf{Deep Encoder $\Phi$}, which encodes the query $q$ into a $\mathbb{D}$-dimensional dense embeddding. Following the previous work, BERT~\cite{devlin2018bert} is employed as the encoder.
    \item \textbf{Cluster Nodes}, which consist of root node, intermediate nodes, and leaf nodes. The leaf nodes correspond to the fine-grained clusters, while the intermediate nodes correspond to the coarse-grained clusters. 
    All clustering nodes are represented as trainable dense embeddings $ \widetilde{e}_{c_k} \in \mathbb{R}^ \mathbb{D}$, where $k$ denotes the $k$-th clustering node.
    More specifically, given the dense embedding of a query $\Phi(q) \in \mathbb{R}^\mathbb{D}$, the relevance scores of the query and the clustered nodes are calculated by $s=\widetilde{e}_{c}^T \cdot \Phi(q)$ at each level of the tree. Based on these scores, top $b$ leaf nodes are returned and the documents within them are further ranked. The parameter $b$ represents the beam size. Cluster node embeddings are crucial for a tree. We will describe how to optimize them in the following section.
    
    \item \textbf{Cluster Assignment}, which represents the distribution of documents to leaf nodes. Assume there are $\mathbf{K}$ leaf nodes, we use $S^{i} = \{ d_1^i,d_2^i,...\}$ to denote the documents assigned to leaf node $i$. The initial cluster assignment is constructed by the k-means algorithm on document embeddings $ \widetilde{e}_{d_k} \in \mathbb{R}^\mathbb{D}$, where $k$ denotes the $k$-th document.
    The score $s=\widetilde{e}_{d}^T \cdot \Phi(q)$ is applied to indicate the relevance of the query to the document. As mentioned above, we only calculate the relevance score for the documents in the top $b$ leaf nodes. Since k-means is an unsupervised clustering method, which divides documents into mutually exclusive clusters where documents in each cluster share the same semantics. However, this does not correspond to reality. Therefore, we relax this constraint and apply an overlapped cluster approach to optimize cluster assignment. The details will be described in section ~\ref{overlap}.
   
\end{itemize}

In this paper, we define $\beta$ as the branch balance factor, representing the number of child nodes of non-leaf nodes, and $\gamma$ as the leaf balance factor, representing the maximum number of documents contained by one leaf node.

JTR builds the tree index by recursively using the k-means algorithm. Specifically, given the corpus to be indexed, all documents are encoded into embeddings with the trained document encoder. Then, all embeddings are clustered into $\beta$ clusters by the k-means algorithm. For each node that contains more than $\gamma$ documents, the k-means is applied recursively until all nodes contain less than $\gamma$ documents. The embedding of each node is initialized as the cluster centroid embedding. Figure ~\ref{cluster} illustrates the initialization of the tree structure. We can observe that the tree index has the following properties: (1) Each non-leaf node of the tree has $\beta$ child nodes. The depth of the tree is influenced by branch balance factor $\beta$ and leaf balance factor $\gamma$. (2) The tree may be unbalanced. In the large corpus, this unbalance is usually insignificant and does not significantly affect the effectiveness of retrieval.

\begin{figure}[t]
\subfigure[Workflow of existing tree-based methods]{
\label{workflow1}
\centerline{\includegraphics[width=0.8\columnwidth]{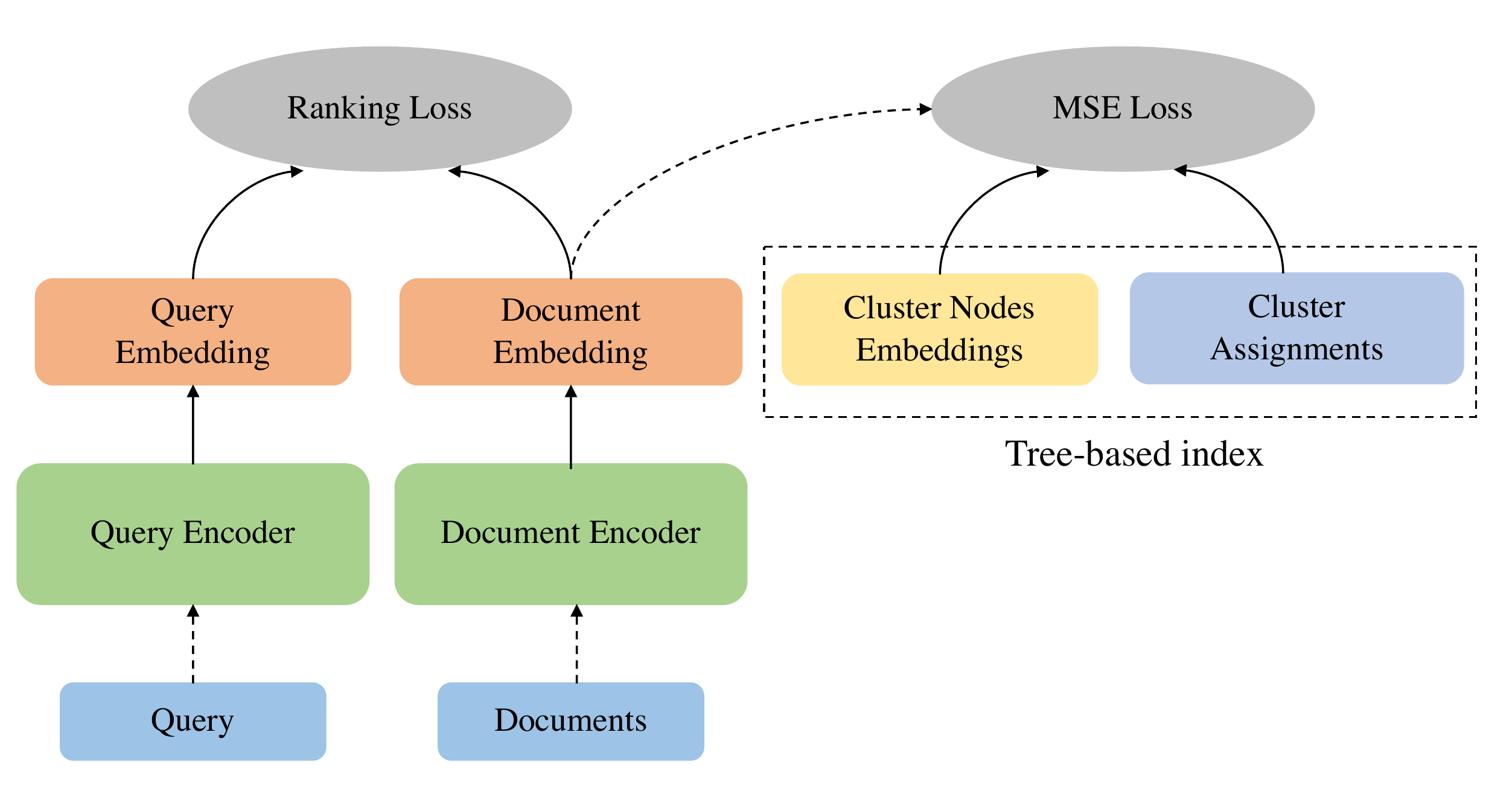}}
}
\subfigure[Workflow of JTR]{
\label{workflow2}
\centerline{\includegraphics[width=0.8\columnwidth]{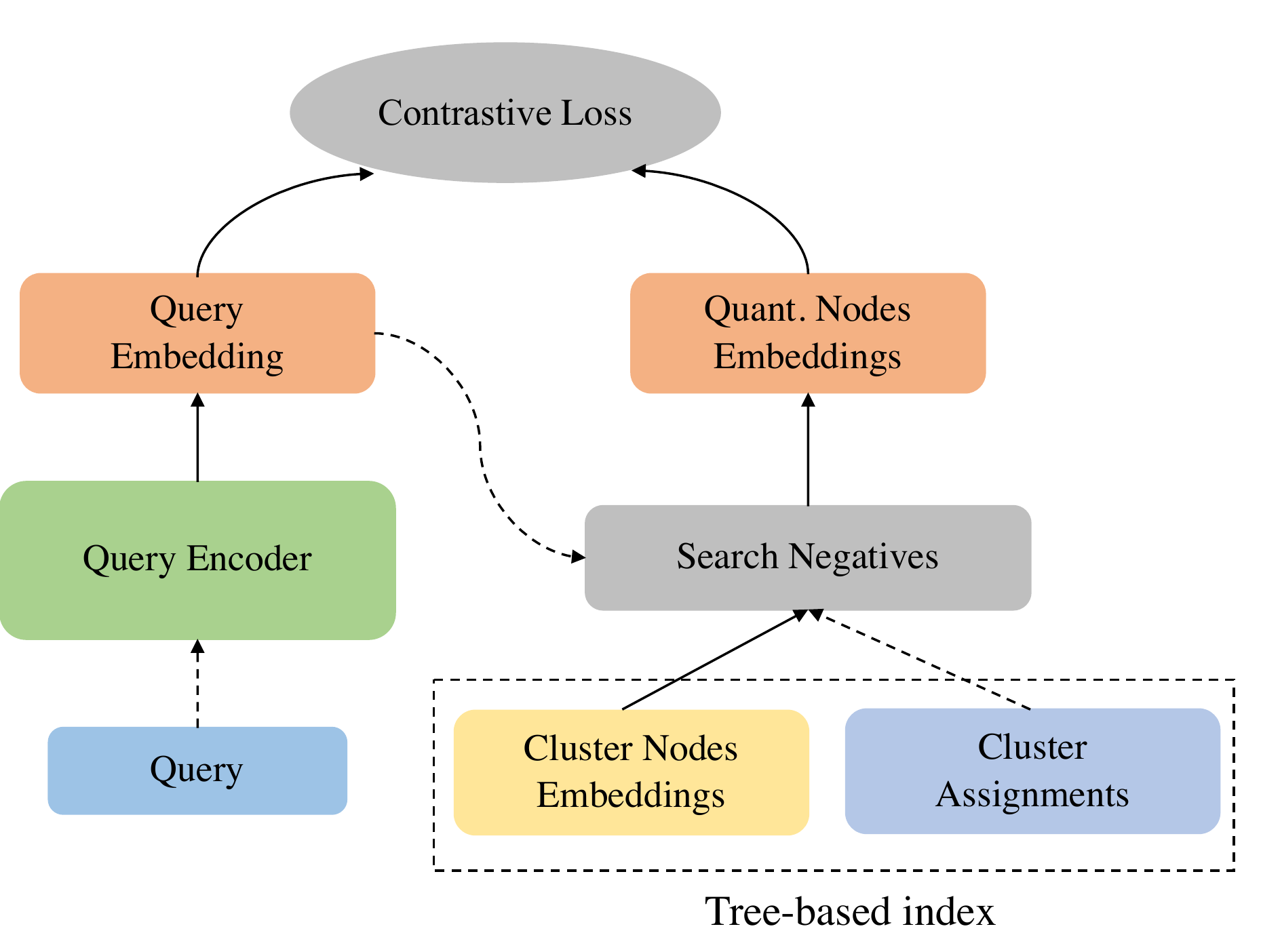}}
}
\vspace{-0.15in}
\caption{Comparison of the workflow of JTR and existing methods. The solid arrows indicate that the gradient propagates backward, while the dashed arrows indicate that the gradient does not propagate.}
\vspace{-5mm}
\label{workflow}
\end{figure}

\subsection{End-to-End Optimization}
In this section, we introduce the end-to-end joint optimization process. In Figure ~\ref{workflow}, we compare the workflow of JTR and existing works. As shown in Figure ~\ref{workflow1}, the existing methods follow a two-step ``encoding-indexing" process. They first train the query encoder and document encoder with the ranking loss. After that, the well-trained document embeddings are used to train the tree-based index under the guidance of MSE loss. The training process of tree-based index is independent and cannot benefit from supervised data.
In contrast, Figure ~\ref{workflow2} shows the optimization process of JTR. The joint optimization is mainly implemented by two strategies: unified contrastive learning loss and tree-based negative sampling. Next, we describe our motivation and the specific design in detail.

\subsubsection{Motivation}

For tree-based indexes, ensuring retrieval effectiveness and efficiency relies on two main components: pruning low-quality nodes and cluster assignment. 
For pruning low-quality nodes, tree-based indexes usually utilize beam search~\cite{beamsearch} to achieve efficient top $b$ leaf nodes. Pruning the correct node at the upper levels can lead to a serious accumulation of errors. From this perspective, we argue that tree-based indexes should have the maximum heap property, which can well support the effectiveness of beam search.
More specifically, the maximum heap property can be expressed as follows:

\begin{equation}\label{eqn-2} 
  p^j(n|q)={\frac{\underset{n_c \in \{ n^{'}s\,children\,node\,in\,level\, j+1\}}{max}p^{j+1}(n_c|q)}{\alpha^{(j)}}}
\end{equation}
Where $p^j(n|q)$ represents the relevant probability between query $q$ and cluster node $n$ in level $j$ and ${\alpha^{(j)}}$ is the normalized term in level $j$. This formula indicates that the relevant probability of a parent node relies on the maximum relevant probability of its children nodes. In other words, given a specific query $q$, the parent of the optimal top node also belongs to the top node of the upper level. The maximum heap property is the basis of beam search~\cite{beamsearch}.

In practice, we do not need to know the exact relevant probabilities of the cluster nodes. The relevance rank order of the nodes at each level is sufficient for accurate beam search. Therefore, we design a new contrastive learning loss to optimize cluster nodes. Moreover, the tree-based negative sample sampling technique is applied to improve the performance of JTR.

\subsubsection{Unified Contrastive  Learning Loss}
Existing DR models usually use the contrastive learning loss function. Specifically, given a query $q$, let $d^+$ and $d^-$ be relevant documents and negative documents. The loss function is formulated as follows: 
\begin{equation}\label{eqn-22} 
  L(q,d^+,d^-_{1},......,d^-_{n}) =
-\log_{}{    \frac{exp(s(q,d^+))}{exp(s(q,d^+))+\sum_{j=1}^nexp(s(q,d^-_j))}}
\end{equation}
Where $s(,)$ is the relevance score.
The purpose of contrastive learning loss is to make the query closer to related documents in the embedding space compared to the irrelevant ones. However, this loss function optimizes the embeddings of queries and documents and is not applicable to existing ANN methods. In JTR, we adapt this loss to optimize the tree-based index. Specifically, given a training data $(q_k,d_k)$, $n_k$ is the leaf node that contains $d_k$. Therefore, $n_k$ is the positive sample of the current level, and the leaf nodes not containing $d_k$ are negative samples. 
On this basis, the ancestor nodes of $n_k$ are also treated as positive samples of the level in which they are located. To make the tree-based index meet the properties of the maximum heap, we optimize it with negative sampling at each level. Let $n^+$ denotes the node embedding of positive samples and $n^-$ denotes the node embedding of negative samples. The unified contrastive learning loss is formalized as follows:

\begin{equation}\label{eqn-3} 
  L(q,n^+,n^-_{1},......,n^-_{n}) =
-\log_{}{    \frac{exp(s(q,n^+))}{exp(s(q,n^+))+\sum_{j=1}^nexp(s(q,n^-_j))}}
\end{equation}
where $s(,)$ is the inner product. The unified contrastive learning loss function jointly optimizes the query encoder parameters and the embedding of clustering nodes. Since the ancestor node of the positive sample remains the positive sample during training, the tree-based index can meet the maximum heap property under the guidance of this loss function.

Training the tree-based index with the contrastive learning loss is non-trivial because the clustering assignment is not differentiable. To solve this problem, we initialize the cluster assignments using Figure~\ref{workflow1} and only train the cluster node embeddings, which can benefit from supervised data directly.

\begin{figure}[t]
\centerline{\includegraphics[width=0.9\columnwidth]{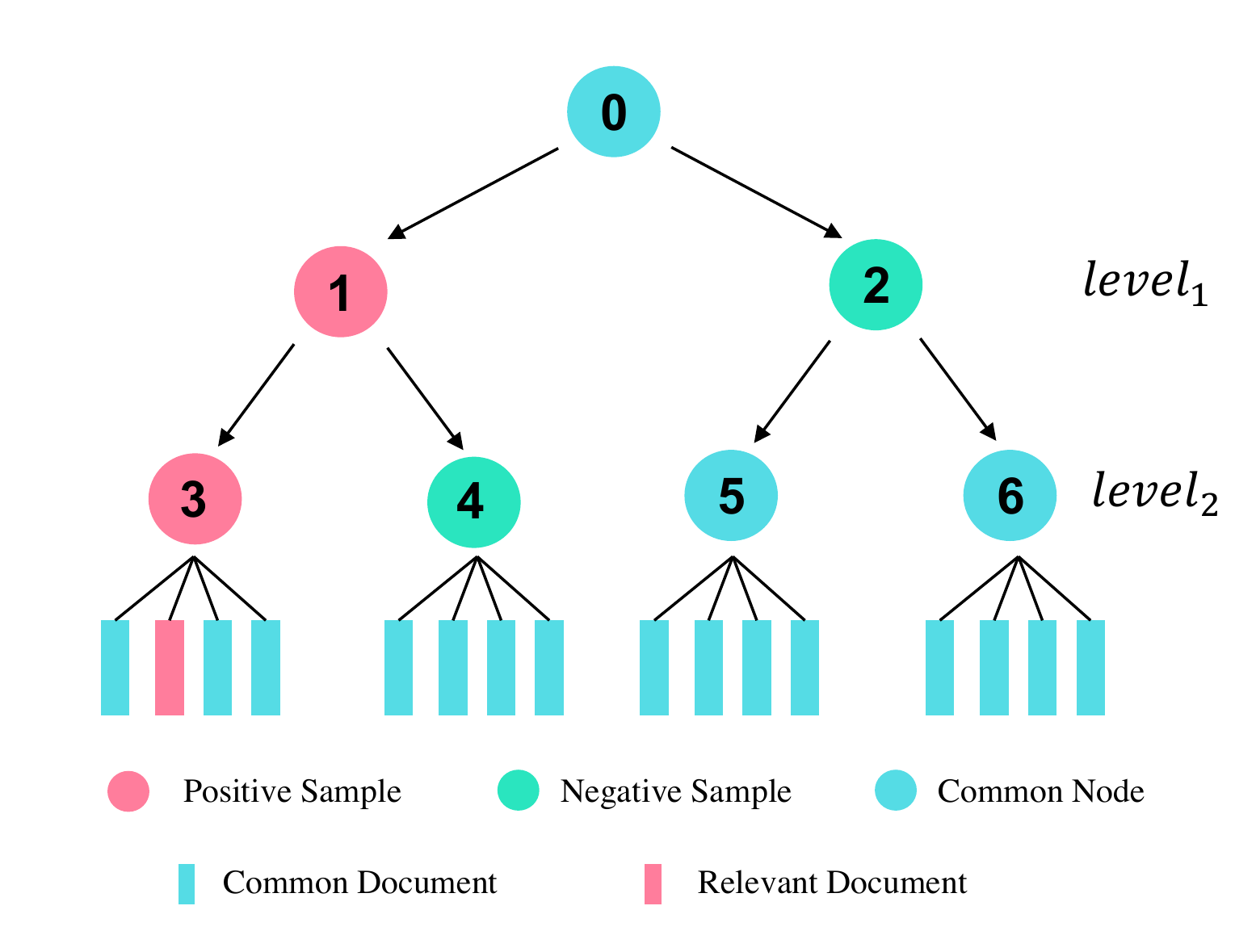}}
\vspace{-0.15in}
\caption{The sampling process of JTR. The integer represents the sequence number of the node. We select the brother nodes of positive samples as negative samples.}
\vspace{-5mm}
\label{negative}
\end{figure}
\subsubsection{Tree-based Negative Sampling}
To further improve the performance of JTR, we design the tree-based negative sampling strategy. As shown in Figure ~\ref{negative}, given a specific query, the leaf node with number three is a positive sample since it contains the relevant document. As mentioned above, the leaf node with number one is a positive sample at $level_1$ because it serves as the father node of the positive sample. Since the tree is constructed with the k-means algorithm, the brother nodes of each positive sample are considered to be closer to the positive sample in the embedding space. Hence, we select them as negative samples for training. As shown in Figure ~\ref{workflow2}, the query embedding is fed into the ``Search Negatives" module and brother nodes of the positive sample are returned under the current parameter.

\begin{figure*}[ht]
\vspace{-0.15in}
\centering
\includegraphics[width=0.9\textwidth]{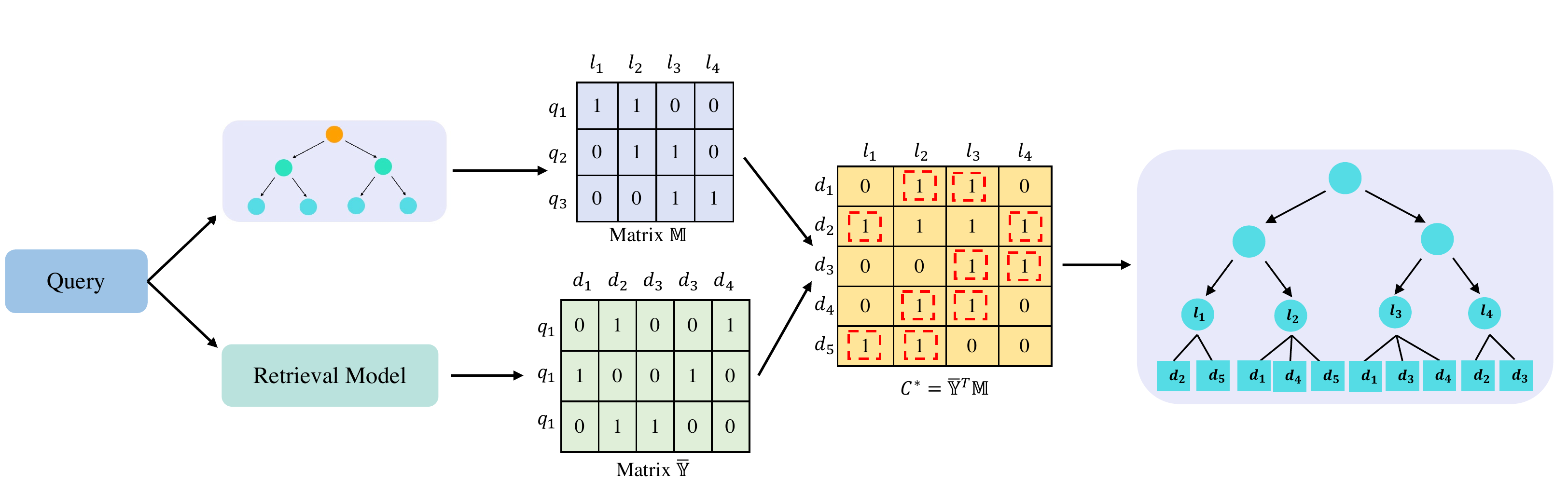}
\vspace{-0.15in}
\caption{Illustration of the optimized overlapped cluster. In this case, there are 3 queries, 4 leaf nodes, and 5 documents. The $q_i \backslash l_i \backslash d_i$ represent the i-th query$\backslash$leaf node$\backslash$document respectively. We set the number of overlapped clustering $\lambda=2$. The values in the red boxes are identified by the $\textit{Proj}(.)$ function. In practice, if a document has the same value for two leaves in C*, the $\textit{Proj}(\cdot)$ function prefers to keep the document in its original leaf.}
\vspace{-5mm}
\label{update}
\end{figure*}

\begin{algorithm}[t]
	\caption{Tree Retrival algorithm}
	\label{Retrival}
	\KwIn{the trained tree $\mathcal{T}$, query $q$, beam size $b$, query encoder $\Phi(\cdot)$}
	\KwOut{$k$ nearest approximate candidates }
    
    Result set $A=\emptyset$, candidate set $Z$=\{the root node $n_1$\}.
    
    \While{$Z \not =\emptyset$}{
    Remove all leaf nodes from $Z$ and insert them into $A$.
    
    Calculate $s=\widetilde{e}_{c_n}^T \cdot \Phi(q)$ for each remaining node $n\in Z$.  ${e}_{c_n}$ is the embedding of current node.
    
    According to the $s$, top $(b-len(A))$ nodes in $Z$ are selected to form the set $I$. There are no leaf nodes in $I$.
    
    Update the $Z$: $Z$ = \{children nodes of $n|n \in I$\}.

    }
    Compute $s=\widetilde{e}_{d}^T \cdot \Phi(q)$ for documents $d$ contained by nodes in $A$ to get top $k$ candidates.
\end{algorithm}

\subsection{Optimized Overlapped Cluster}
\label{overlap}
As mentioned above, cluster assignment plays an essential role in the tree-based index.
However, k-means is an unsupervised clustering method, which divides documents into mutually exclusive clusters, and documents in each cluster should share the same semantics.  We observe in practice that the semantics of a document are complex and multi-topic. Putting each document in a cluster can limit the performance of the tree-based index. To solve this problem, we propose overlapped clusters to further optimize cluster assignment.

The overlapped cluster has been studied in the field of unsupervised learning~\cite{cleuziou2008extended,whang2015non,liu2021label}. Inspired by Liu et al. ~\cite{liu2021label}, we formulate the cluster assignment problem in information retrieval as an optimization problem.
Suppose that there are $\mathbf{L}$ queries and $\mathbf{N}$ documents in the training corpus and all documents have been clustered and put into the $\mathbf{K}$ leaf nodes of the tree index.

First, we define the ground truth matrix $\mathbb{Y} \in \{0,1\}^{\mathbf{L} \times \mathbf{N}} $ as:

\begin{equation}\label{eqn-24} 
  	\mathbb{Y}_{i,j}=\left\{
		\begin{aligned}
			1 & & \text{if\,$doc_j$\,is relevant to\,\,$query_i$}  \\
			0 & & \text{otherwise}  
		\end{aligned}
		\right.
\end{equation}
$\mathbb{Y}$ describes the relevance of the training queries and the documents and is the best result the model is trying to achieve.

Then, we input the training queries into the tree-based index to get the leaf nodes associated with them. We define the relationship between queries and leaf nodes as matrix 
$\mathbb{M} \in \{0,1\}^{\mathbf{L} \times \mathbf{K}} $ and the original cluster assignment as matrix $\mathbb{C} \in \{0,1\}^{\mathbf{N} \times \mathbf{K}}$. It is worth noting that the initialized cluster assignment is obtained from Figure ~\ref{workflow1}.
\begin{equation}\label{eqn-4} 
  	\mathbb{M}_{i,j}=\left\{
		\begin{aligned}
			1 & & \text{if\,$query_i$\,matches\,$leaf_j$}  \\
			0 & & \text{otherwise}  
		\end{aligned}
		\right.
\end{equation}

\begin{equation}\label{eqn-5} 
	\mathbb{C}_{i,j}=\left\{
		\begin{aligned}
			1 & & \text{if\,$doc_i$\,matches\,\,$leaf_j$}  \\
			0 & & \text{otherwise}  
		\end{aligned}
		\right.
\end{equation}
Then we can represent the relationship between queries and documents of JTR as:
\begin{equation}\label{eqn-6} 
  \hat{\mathbb{Y}} = \textit{Binary}(\mathbb{M} \times \mathbb{C}^\mathrm{T})
\end{equation}
where $\textit{Binary}(A) = I_A$ is the element-wise indicator function. When the element $a>0$, $I_a=1$, which ensures $\hat{\mathbb{Y}} \in \{0,1\}^{\mathbf{L} \times \mathbf{N}}$. $\times$ stands for matrix cross product. Then, the performance of JTR can be expressed as the intersection of $\mathbb{Y}$ and $\hat{\mathbb{Y}}$.
The formula is as follows:
\begin{equation}\label{eqn-7} 
  \textit{Recall} = |\hat{\mathbb{Y}} \cap \mathbb{Y}|
\end{equation}
where $|\cdot|$ returns the number of non-zero elements in the matrix. Since $\hat{\mathbb{Y}}$ and $\mathbb{Y}$ are binary matrices, We can obtain the following formula:
\begin{equation}\label{eqn-8} 
  \textit{Recall} = |\hat{\mathbb{Y}} \cap \mathbb{Y}| = Tr(\mathbb{Y}^\mathrm{T} \times \hat{\mathbb{Y}})
\end{equation}
where $Tr(\cdot)$ returns the trace of the matrix. It is found that the performance of JTR is proportional to the trace of the matrix. In order to maximize the recall rate, we formulate the cluster assignment problem as follows:
\begin{equation}\label{eqn-9} 
  \underset{S_1,S_2,......,S_k}{\textit{maximize}}Tr(\mathbb{Y}^\mathrm{T} \times \textit{\textit{Binary}}(\mathbb{M} \times \mathbb{C}^\mathrm{T})
\end{equation}
\begin{equation}\label{eqn-11} 
 s.t. \sum_{i=1}^K I_{d \in S_i}\leq \lambda, \forall d \in \{d_1,......d_N\}
\end{equation}
\begin{equation}\label{eqn-12} 
  \bigcup_{i=1}^K S_i = \{d_1,......d_N\}
\end{equation}
where $S_i$ represents the $i$ th cluster assignment. $I$ is the indicator function. Equation ~\ref{eqn-11} guarantees that any document $d$ appears at most $\lambda$ times in all clusters. Equation ~\ref{eqn-12} ensures that the clusters include all documents.

This optimization problem is an NP-complete problem. Following Liu et al.~\cite{liu2021label}, we approximate the objective function with a continuous, RelU-like function.
\begin{equation}\label{eqn-15} 
  \textit{Binary}(\mathbb{M} \times \mathbb{C}^\mathrm{T}) \approx max(\mathbb{M} \times \mathbb{C}^\mathrm{T},0) = \mathbb{M} \times \mathbb{C}^\mathrm{T}
\end{equation}

In the optimization function, the cluster information is hidden in the matrix $\mathbb{C}$. $Tr(\mathbb{Y}^\mathrm{T} \times \mathbb{M} \times \mathbb{C}^\mathrm{T})$ is linear with matrix $\mathbb{C}$, so the optimal solution of $\mathbb{C}$ is the projection of $\mathbb{Y}^\mathrm{T} \times \mathbb{M}$ onto the constraint set. Therefore, the optimization problem has a closed-form solution:
\begin{equation}\label{eqn-16} 
  C^\ast = \textit{Proj} ( \mathbb{Y}^\mathrm{T} \times \mathbb{M} ) 
\end{equation}
where the $\textit{Proj}(\cdot)$ operator selects the top $\lambda$ elements for each line of the matrix. In the field of Information Retrieval, there exists the problem of data sparsity. Only a small set of the documents have the corresponding training queries, resulting in a lot of rows in $\mathbb{Y}^\mathrm{T}$ with all entities being zeros. To solve this problem, we use the trained DR model to retrieve the top-$k$ documents of each training query to construct matrix $\bar{\mathbb{Y}} \in \{0,1\}^{\mathbf{L} \times \mathbf{N}}$. In other words, if $doc_j$ is the top $k$ document recalled by $query_i$ using the DR model, then $\bar{\mathbb{Y}}_{i,j} = 1$. Hence, every line of $\bar{\mathbb{Y}}$ has $k$ nonzero entries. In short, the final solution is:
\begin{equation}\label{eqn-17} 
  C^\ast = \textit{Proj} (\bar{\mathbb{Y}}^\mathrm{T} \times \mathbb{M})
\end{equation}

The proposed solution is based on the intuition that two documents are more likely to cluster into the same class if they are related to the same query. It is worth noting that only the cluster assignment is changed in this process, the structure of the tree and the cluster node embeddings do not change.

After the $\mathbb{C}^\ast$ is determined, We re-optimize the JTR as described in Section 3.2 to accommodate the new cluster assignment. Figure \ref{update} shows the process of optimized overlapped cluster. We acknowledge that there exists a few documents that are not relevant to any training query. These documents are retained in the original clusters.


\begin{table*}[ht]
  \caption{
  Results on the MS MARCO dataset. AQT stands for Average Query processing Time. which is measured by averaging time over each query of the MS MARCO Dev set with a single thread and a single batch on the CPU. */** denotes that JTR performs significantly better than baselines at $p < 0.05/0.01$ level using the two-tailed pairwise t-test. The best method in each column is marked in bold. 
  }
  \vspace{-4mm}
  \label{table1}
  \begin{tabular}{l|lllll|lllll}
  \hline
  \multicolumn{1}{c|}{\multirow{2}{*}{\textbf{Model}}} & \multicolumn{2}{l}{\textbf{MARCO Passage}} & \textbf{DL19 Passage} & \textbf{DL20 Passage} & \textbf{AQT} & \multicolumn{2}{l}{\textbf{MARCO Doc}} & \textbf{DL19 Doc} & \textbf{DL20 Doc} & \textbf{AQT} \\
  \multicolumn{1}{c|}{}                                & MRR@100              & R@100               & NDCG@10               & NDCG@10               & ms           & MRR@100            & R@100             & NDCG@10           & NDCG@10           & ms           \\ \hline
  IVFFlat                                              & 0.311*               & 0.778               & 0.580*                & 0.615                 & 75           & 0.349**            & 0.839             & 0.572             & 0.527*            & 63           \\
  PQ                                                   & 0.289**              & 0.717**             & 0.448**               & 0.546**               & 149          & 0.290**            & 0.788**           & 0.498**           & 0.458**           & 58           \\
  IVFPQ                                                & 0.252**              & 0.653**             & 0.532**               & 0.540**               & 15           & 0.279**            & 0.695**           & 0.465**           & 0.423**           & 10           \\
  JPQ                                                  & 0.306**              & \textbf{0.832}      & \textbf{0.611}        & 0.607                 & 152          & 0.347**            & \textbf{0.889}    & 0.575             & 0.536             & 55           \\
  Annoy                                                & 0.144**              & 0.263**             & 0.361**               & 0.385**               & 132          & 0.148**            & 0.253**           & 0.504**           & 0.463**           & 57           \\
  FALCONN                                              & 0.295**              & 0.719**             & 0.554**               & 0.532**               & 23           & 0.321**            & 0.756**           & 0.496**           & 0.460**           & 9            \\
  FLANN                                                & 0.271**              & 0.629**             & 0.551**               & 0.578**               & 18           & 0.294**            & 0.649**           & 0.418**           & 0.368**           & 5            \\
  IMI                                                  & 0.314                & 0.697**             & 0.542**               & 0.565**               & 37           & 0.348**            & 0.788**           & 0.535*            & 0.468**           & 26           \\
  HNSW                                                 & 0.289**              & 0.732**             & 0.546**               & 0.559**               & 11           & 0.334**            & 0.783**           & 0.503**           & 0.507**           & 5            \\
  JTR                                                  & \textbf{0.318}       & 0.778               & 0.610                 & \textbf{0.632}        & 30           & \textbf{0.364}     & 0.848             & \textbf{0.590}    & \textbf{0.565}    & 18           \\ \hline
  \end{tabular}
  \vspace{-4mm}
  \end{table*}

\subsection{Tree Retrival}
For a specific query, the tree retrieval process is described in algorithm \ref{Retrival}. At each level of the tree, we select the top nodes and their child nodes as candidates for the next level. It saves retrieval time by pruning less relevant nodes. In the tree structure of JTR, the leaf nodes are not always at the same level. Therefore, the $A$ set is used to preserve the encountered leaf nodes. In line 5 of Algorithm \ref{Retrival}, we select top $b-len(A)$ nodes in $Z$ each time to create $I$. Since the leaf nodes in $Z$ are removed and inserted into $A$ in line 3, there are no leaf nodes in $I$.

In the JTR, the retrieval process is hierarchical and top-to-down. If the tree has $\mathbf{K}$ leaf nodes, $\mathbf{N}$ documents and beam size is $b$ and branch balance factor is $\beta$, the time complexity of retrieving leaf nodes is $O(\beta*b* \log \mathbf{K})$. For retrieval in set $A$, the maximum number of documents in $A$ is $b*\gamma$, where $\gamma$ is leaf balance factor. Since $\gamma$ can be approximated as $\mathbf{N} / \mathbf{K}$, the time complexity of this part is $O(b*\mathbf{N} / \mathbf{K})$. In summary, the retrieval time complexity of JTR is $O(\beta*b* \log \mathbf{K}) + O(b * \mathbf{N} / \mathbf{K})$. The overall time complexity is below the linear time complexity, which greatly improves the retrieval efficiency.

\section{Experiment settings}
In this section, we introduce our experimental settings, including implementation details, datasets and metrics, baselines.

\subsection{Datasets and Metrics}

The experiments are conducted on the dataset MS MARCO~\cite{MARCO}, which is a large-scale ad-hoc retrieval benchmark. It contains two large-scale tasks: Document Retrieval and Passage Retrieval.
Passage Retrieval has a corpus of $8.8M$ passages, $0.5M$ training queries, and $7k$ development queries. Document Retrieval has a corpus of $3.3M$ documents, $0.4M$ training queries, and $5k$ development queries.

We conduct our experiments on two tasks from MS MARCO Dev, TREC2019 DL~\cite{TREC19} and TREC2020 DL~\cite{TREC20}. The MS MARCO Dev is extracted from Bing's search logs, and each query is marked as relevant to a few documents. TREC2019 DL and TREC2020 DL are collections that contain extensively annotated documents for each query, i.e., using four-level annotation criteria. We use $R@100$ to evaluate the recall performance of different methods. $MRR@100$ and $NDCG@10$ are applied to measure ranking performance.

\subsection{Baselines}
We select several state-of-the-art ANN indexes as our baselines. Here are more details:

\textbf{IVFFlat}~\cite{IVF}: IVFFlat is the classic inverted index. When applied to dense retrieval, IVF defines $nlist$ clusters in the vector space, and only the top
$nprobe$ clusters close to the query embedding are retrieved. We set $nprobe$ to be the same as beam size $b$.

\textbf{PQ}~\cite{jegou2010product}: PQ is an ANN index based on product quantization. We set the number of embedding segments to 32 and the number of coding bits to 8.

\textbf{IVFPQ}~\cite{jegou2010product}: IVFPQ is one of the fastest indexes available, which combines ``inverted index + product quantization”. The parameter settings are the same as IVFFlat and PQ.

\textbf{JPQ}~\cite{JPQ}: JPQ implements the joint optimization of query encoder and product quantization. We set the number of segments into which each embedding will be divided as 24.

\textbf{Annoy}~\cite{Annoy}: Annoy improves retrieval efficiency by building a binary tree. We set the number of trees to 100.

\textbf{FALCONN}~\cite{FALCONN}: FALCONN is an optimized LSH method, which supports hyperplane LSH and cross polyhedron LSH. We use the recommended parameter settings as adopted in the corresponding paper.

\textbf{FLANN}~\cite{Flann}: FLANN is one of the most complete ANN open-source libraries, including liner, kdtree, kmeans tree, and other index methods. We use automatic tuning to get the best parameters.

\textbf{IMI}~\cite{IMI}: IMI is a multilevel inverted index, which combines product quantization and inverted index to bring very good search performance. The number of bits is set to 12.

\textbf{HNSW}~\cite{HNSW}: HNSW is a typical and widely-used graph-based index. We set the number of links to 8 and ef-construction to 100.

For a fair comparison, all ANN indexes are operated on the document embeddings formed by STAR~\cite{STAR}. For IVFFlat, PQ, IVFPQ, HNSW and IMI, we implement them with the Faiss~\footnote{\url{https://github.com/facebookresearch/faiss}}. For Annoy~\footnote{\url{https://github.com/spotify/annoy}}, FALCONN~\footnote{\url{https://github.com/FALCONN-LIB/FALCONN}} and FLANN~\footnote{\url{https://github.com/flann-lib/flann}}, we use the official library to implement them.

\subsection{Implementation Details}
We implemented JTR using PyTorch and Transformers~\footnote{\url{https://huggingface.co/docs/transformers/index}}. Initial embeddings were obtained using STAR~\cite{STAR} for all documents. We also load the checkpoint of STAR to warm up the query encoder. In our experiment, 
The default $\gamma$ is 1000 and $\beta$ is 10. The dimension of embeddings is 768. For the training setup, we use the AdamW~\cite{loshchilov2018fixing} optimizer with a batch size of 32. The learning rate is set to $5\times 10^{- 6}$. In the overlapped cluster of the tree, we use the top 100 documents obtained from ADORE-STAR~\cite{STAR} to build matrix $\bar{\mathbb{Y}}$. All experiments are evaluated on the workstation with Xeon Gold 5218 CPUs and RTX 3090 GPUs.

\begin{figure}[t]
\vspace{-3mm}
\subfigure[MRR curves]{
\label{figure1}
\centerline{\includegraphics[width=0.8\columnwidth]{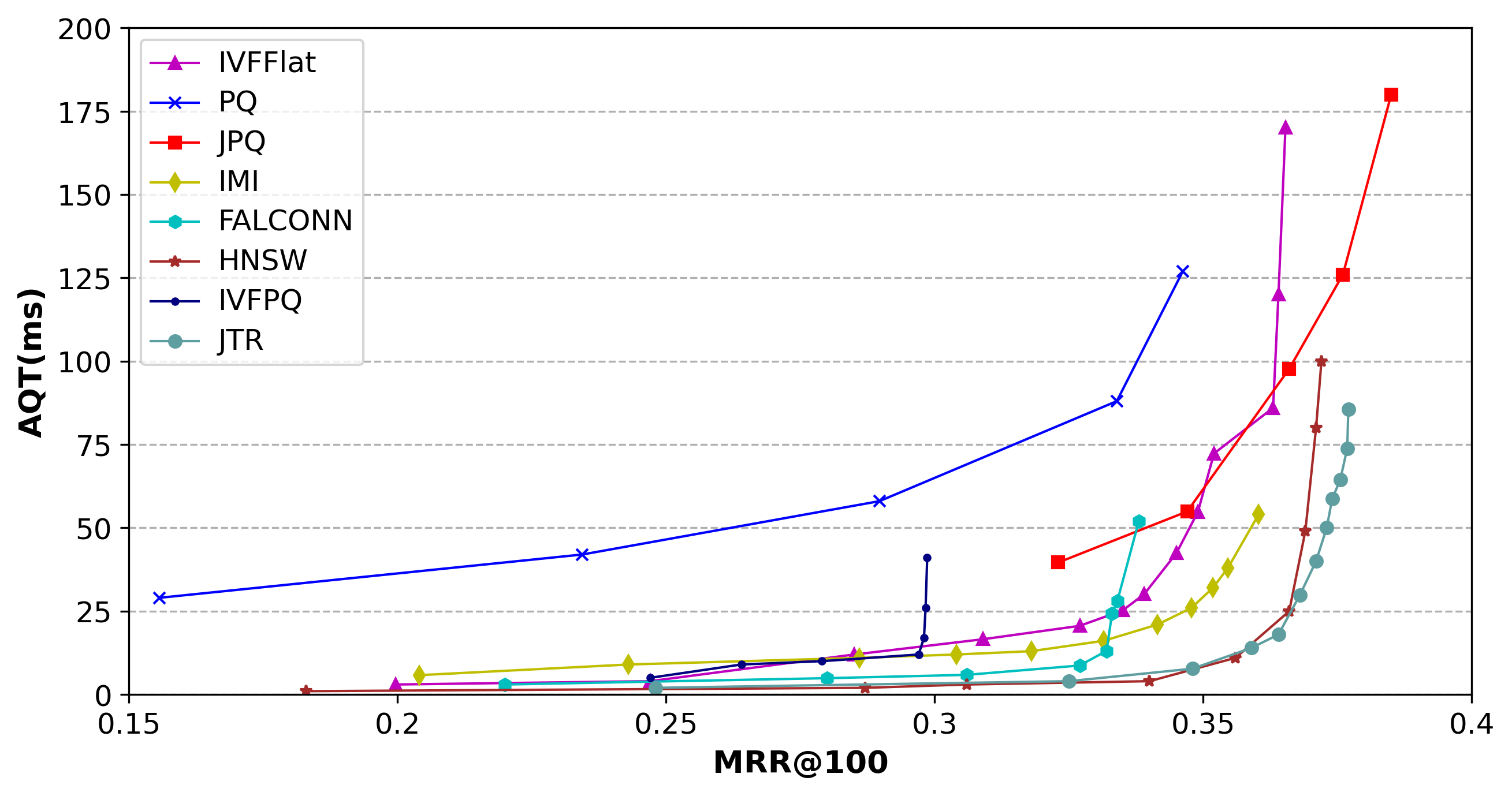}}
}
\subfigure[Recall curves]{
\label{figure2}
\centerline{\includegraphics[width=0.8\columnwidth]{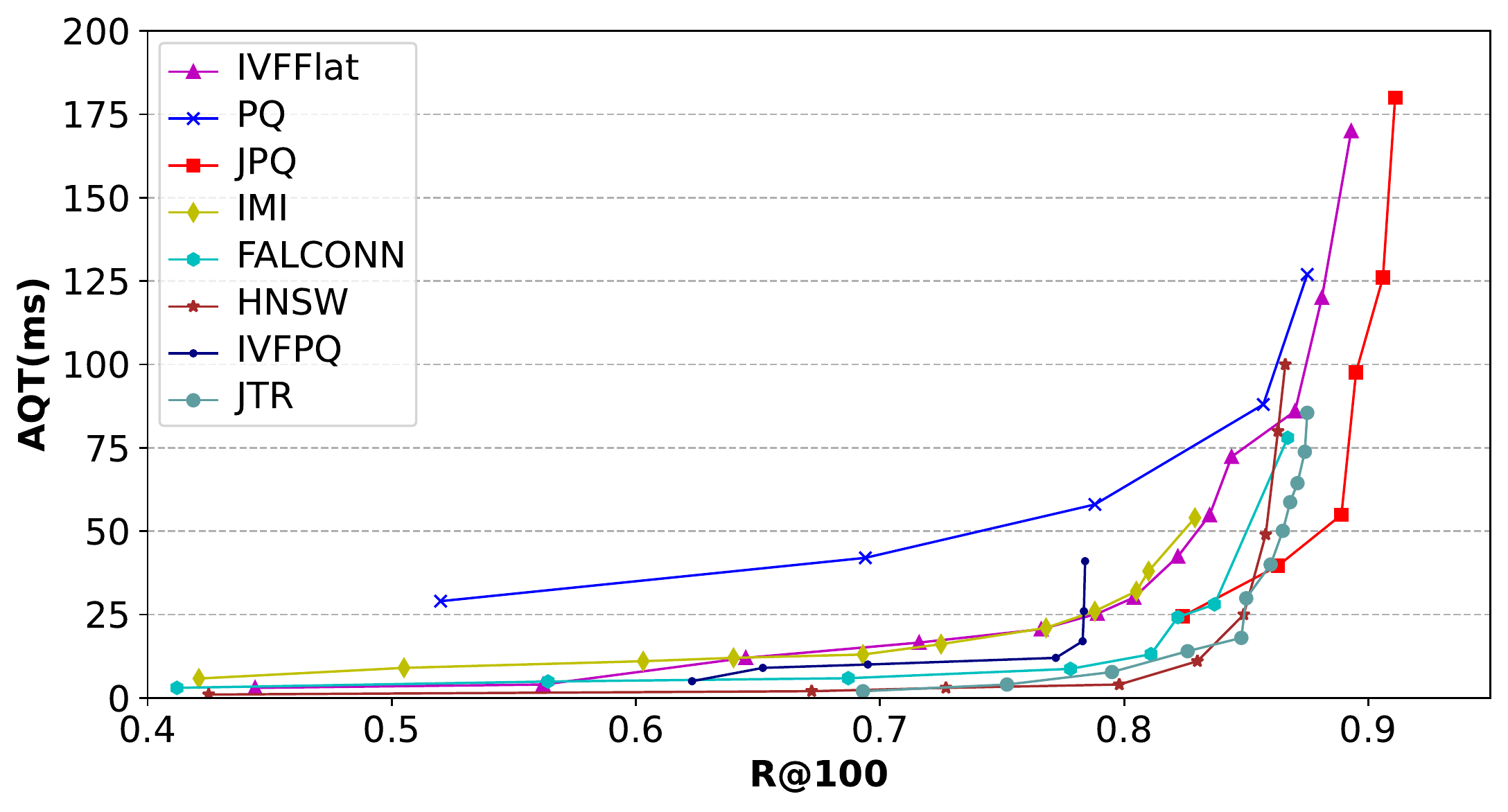}}
}
\vspace{-3mm}
\caption{Trade-off curves for different ANN methods. AQT stands for Average Query processing Time. Bottom and right is better.}
\vspace{-5mm}
\label{tradeoff}
\end{figure}

\section{Experiment results}
\subsection{Comparison with ANN Methods}
The performance comparisons between SAILER and baselines are shown in Table \ref{table1}. We derive the following observations from the experiment results.

\begin{itemize}[leftmargin=*]
    \item PQ and JPQ are ANN methods based on product quantization, which have a linear time complexity with respect to the corpus size. As expected, their average query processing time is the longest among all baselines. IVFPQ is an inverted index based on product quantization, which improves retrieval efficiency but damages retrieval effectiveness.
    \item Benefiting from a different index structure, the average query processing time of IVFPQ, FALCONN, FLANN, and HNSW is lower than that of JTR. However, they suffered a loss in effectiveness.
    \item Annoy and FLANN are existing tree-based indexes. Compared with them, JTR achieved the best results. To the best of our knowledge, JTR is the best tree-based index available, which shows the great potential of tree-bases indexes for dense retrieval.
    \item Overall, JTR performs significantly better than the baseline method on most measures. In terms of effectiveness, JTR achieves the best MRR/NDCG and the second best recall performance (i.e., R@100) on all datasets. Compared to the baseline with the best recall (i.e., JPQ), JTR achieves 3 to 5 times latency speedup. We can conclude that JTR has a very competitive effectiveness compared to other ANN indexes.
\end{itemize}

To further analyze the ability of different indexes to balance efficiency and effectiveness, we plot AQT-MRR curves with varying parameters. As shown in Figure \ref{tradeoff}, we have the following findings:

\begin{itemize}[leftmargin=*]
    \item IVFPQ and FALCONN are limited by resources, which makes it difficult to achieve high retrieval performance.
    \item When our requirement on recall is extremely high (e.g., larger than 0.8), brute-force search could be more efficient than any indexing technique, which makes brute-force search-based algorithms like JPQ more efficient than JTR. After all, any indexing process would add more time complexity if we eventually need to check all documents in the corpus. However, such high requirements on recall are not common in web search or other similar retrieval tasks.
    \item  To sum up, JTR has a very outstanding effectiveness-efficiency trade-off ability, which can always achieve better results with a short retrieval latency. JTR provides a potential solution to balance efficiency and effectiveness for dense retrieval.

\end{itemize}

\begin{figure}[t]
\centerline{\includegraphics[width=0.8\columnwidth]{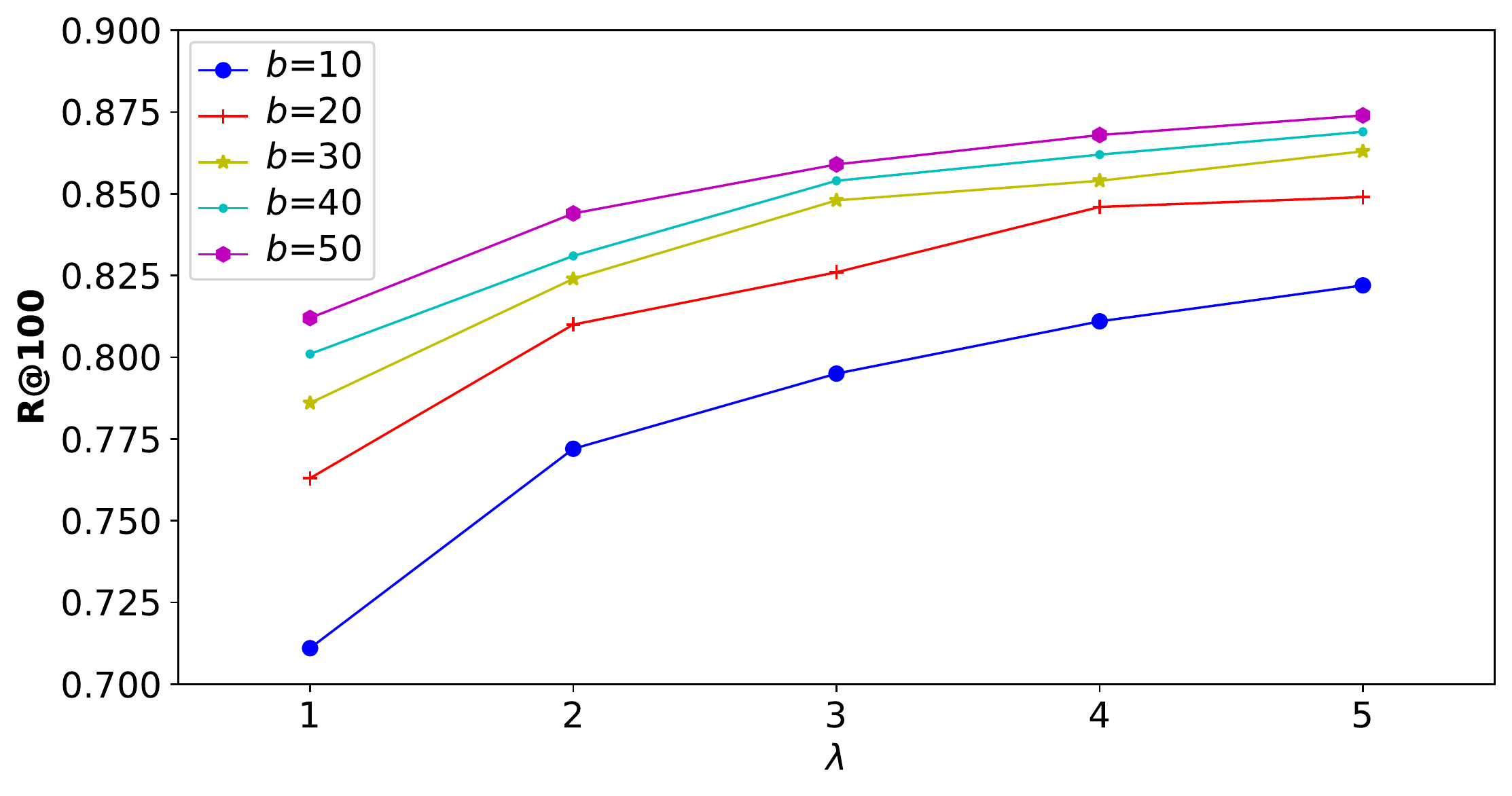}}
\vspace{-0.1in}
\caption{R@100 different $\lambda$ ranging from 1 to 5 and $b$ ranging from 10 to 50.}
\vspace{-5mm}
\label{R}
\end{figure}

\begin{figure}[t]
\centerline{\includegraphics[width=0.8\columnwidth]{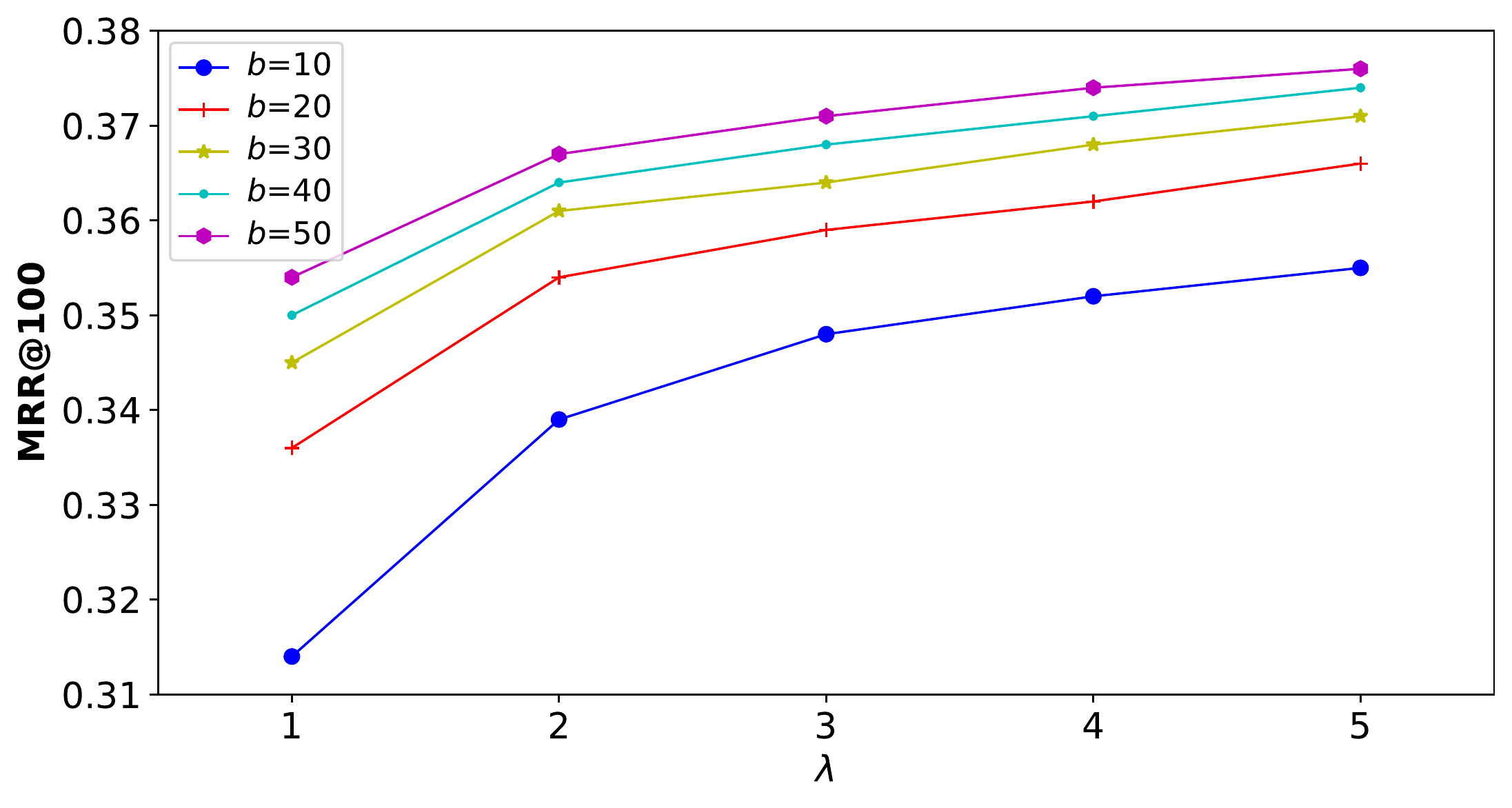}}
\vspace{-0.1in}
\caption{MRR@100 versus different $\lambda$ ranging from 1 to 5 and $b$ ranging from 10 to 50.}
\vspace{-5mm}
\label{MRR}
\end{figure}

\begin{figure}[t]
\centerline{\includegraphics[width=0.8\columnwidth]{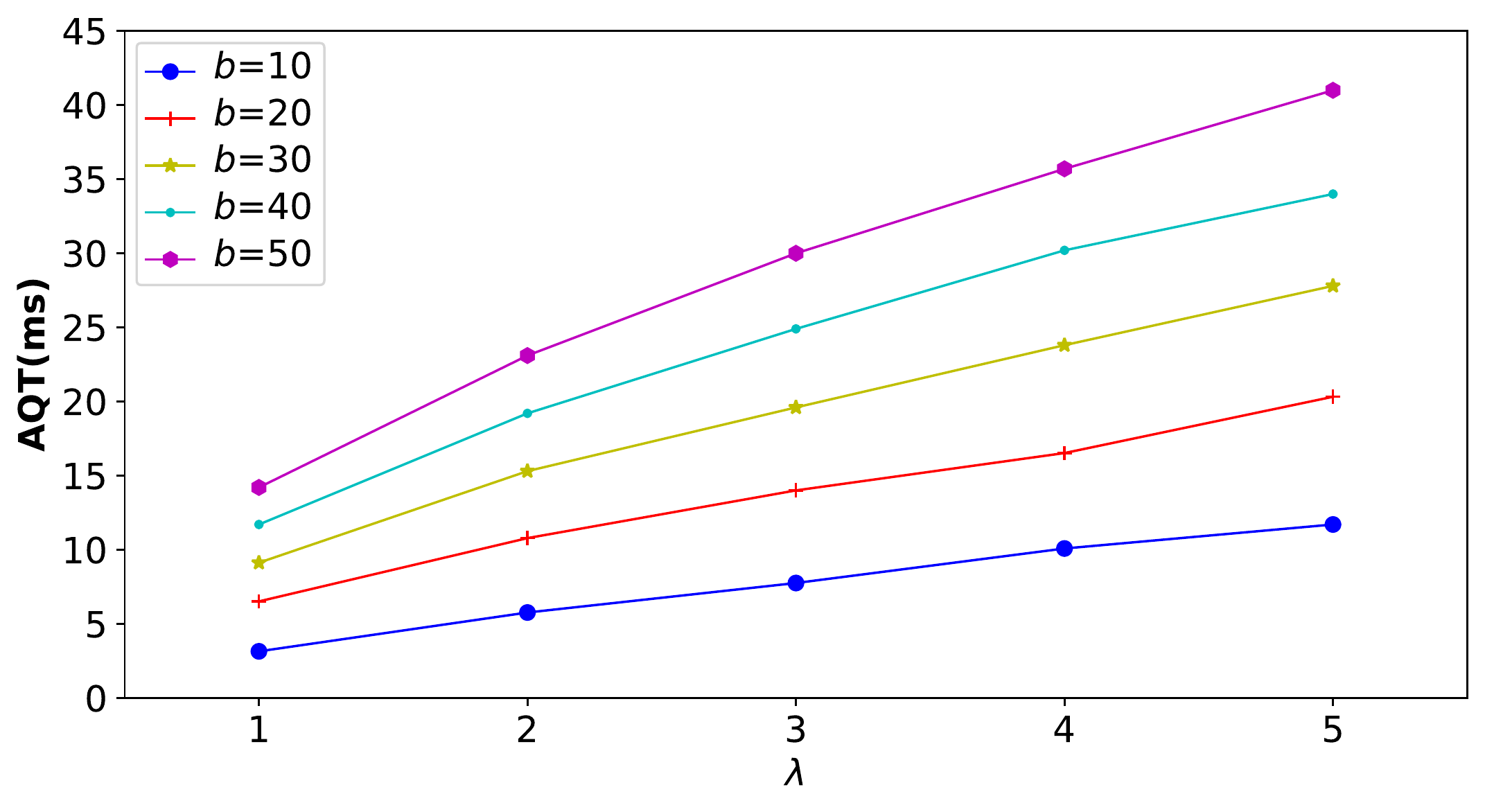}}
\vspace{-0.1in}
\caption{AQT versus different $\lambda$ ranging from 1 to 5 and $b$ ranging from 10 to 50. AQT stands for Average Query processing Time.}
\vspace{-5mm}
\label{AQTs}
\end{figure}

\subsection{Hyperparameter Sensitivity}

\subsubsection{Hyperparameters for Tree Retrieval}
In this section, we investigate the impact of the overlap number $\lambda$ and the beam size $b$. All experiments were conducted on the MS MARCO Doc Dev dataset.

We set $\lambda = 1,2,3,4,5$ and $b = 10,20,30,40,50$ to observe the retrival performance of JTR. Figures \ref{R}, \ref{MRR} and \ref{AQTs} show the changes in retrieval quality and retrieval efficiency with hyperparameters. We can get the following observations:

\begin{itemize}[leftmargin=*]
    \item Both retrieval quality and latency increase with $\lambda$. When $\lambda$ increases, the leaf node contains more documents. The number of candidate documents goes up which leads to better results and higher latency.
    \item The slope of MRR and R curve decreases with the increase of $\lambda$, but the slope of AQT remains unchanged. That means the gain from overlapped cluster is getting smaller.
    \item When we fix $\lambda$, it is found that the gains of MRR and R decreased with the increase of $b$. The reason may be that the most relevant documents have already clustered in the top leaf nodes. 
    \item In Figure \ref{AQTs}, we note that the gap between curves increases when $\lambda$ is larger, which shows that a large $\lambda$ could lead to larger computation costs on each leaf node. 
    \item Overall, the increase of $b$ and $\lambda$ can lead to better performance and longer latency. Therefore, we recommend the moderate value for $\lambda$ and $b$ to maximize the trade-off between effectiveness and efficiency.
\end{itemize}

\subsubsection{Hyperparameter for Overlapped Cluster}

\begin{figure}[ht]
\vspace{-0.15in}
\subfigure[MRR curves]{
\label{figure5}
\centerline{\includegraphics[width=0.8\columnwidth]{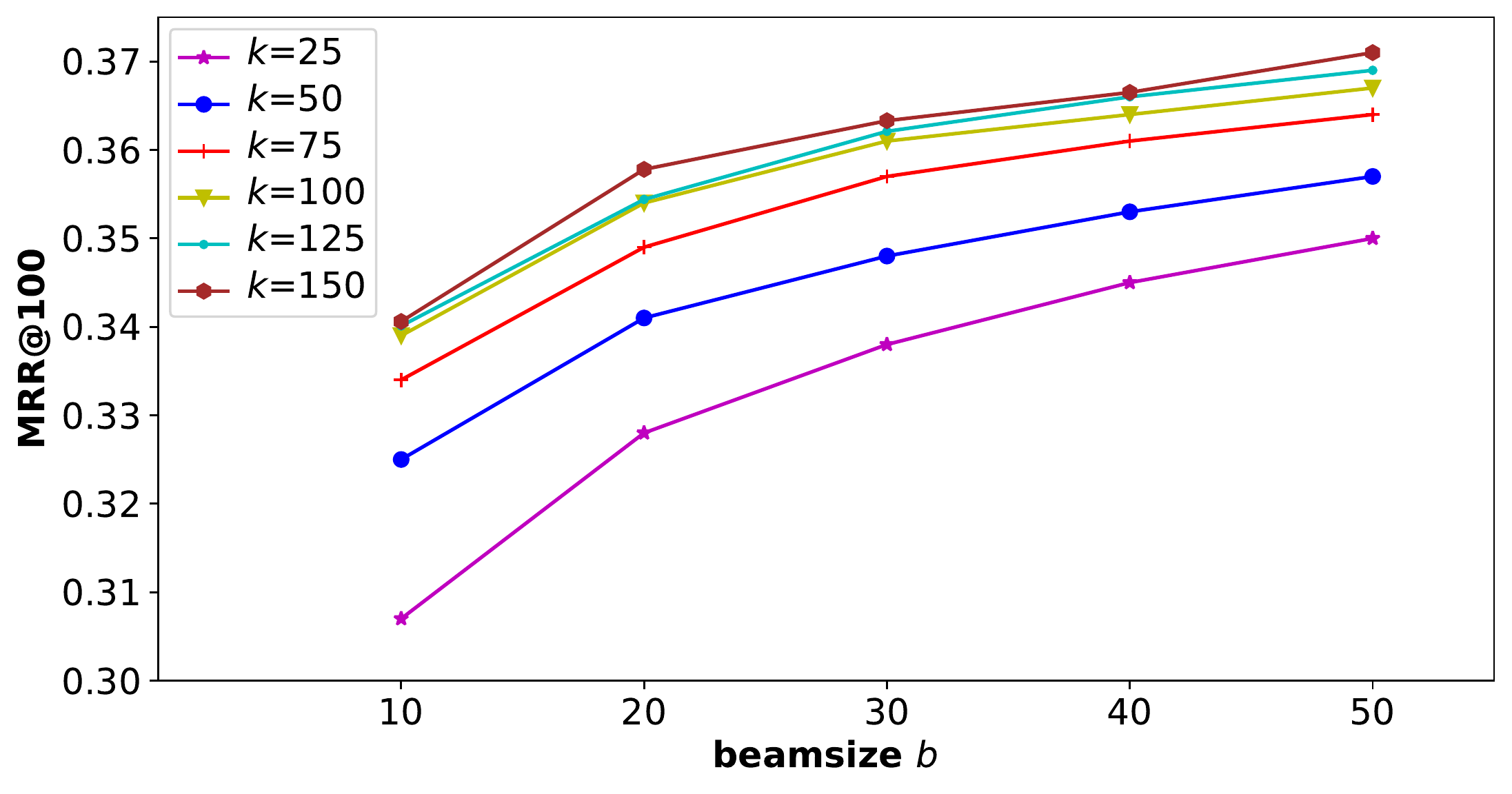}}
}
\subfigure[Recall curves]{
\label{figure6}
\centerline{\includegraphics[width=0.8\columnwidth]{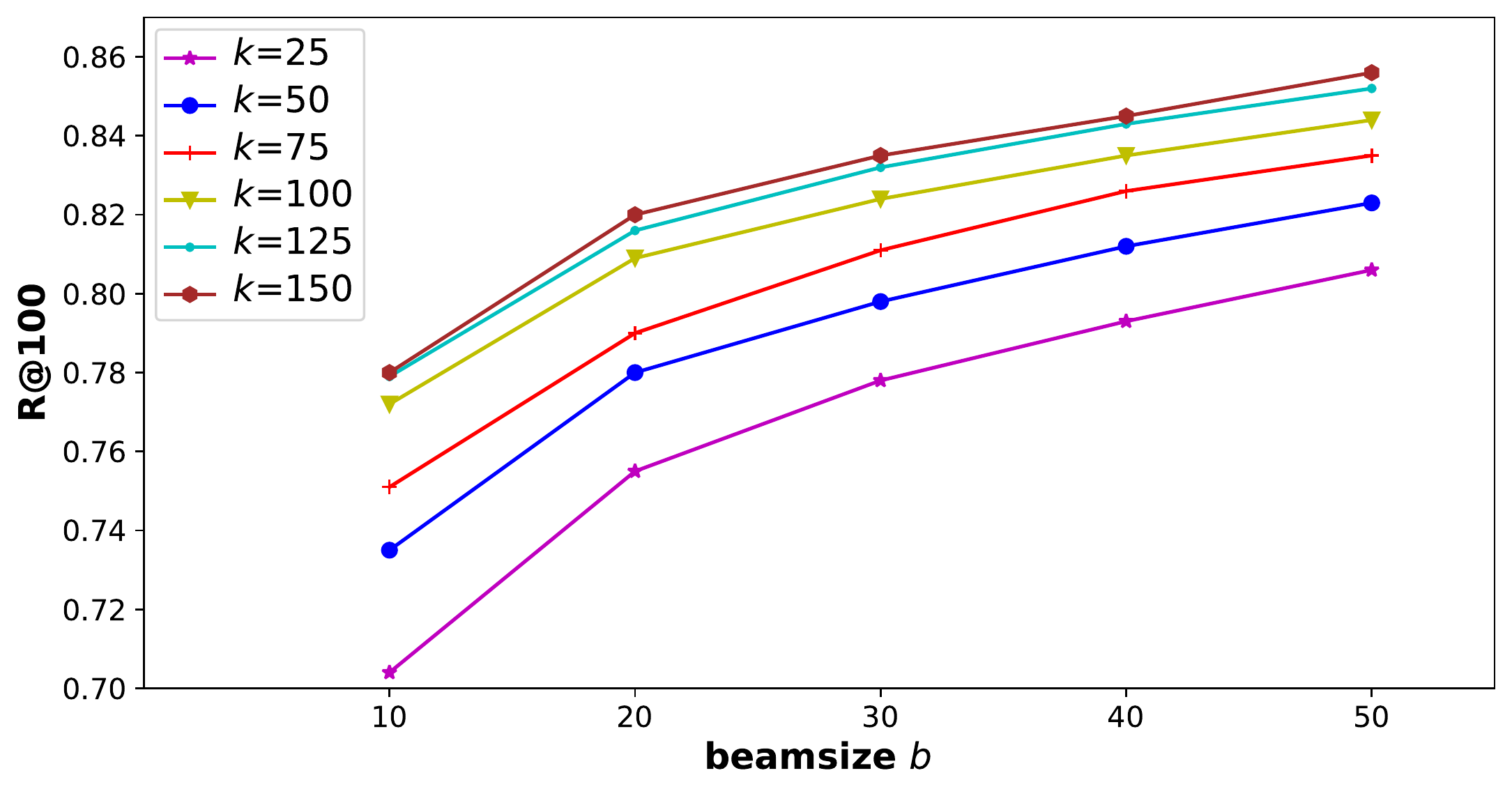}}
}
\vspace{-0.15in}
\caption{MRR@100 and R@100 versus different $k$ ranging from 25 to 150
and $b$ ranging from 10 to 50.}
\vspace{-5mm}
\label{Y}
\end{figure}

In this section, we fix the model as ADORE-STAR and change $k$ from 25 to 150, where $k$ denotes the number of documents recalled with the dense retrieval model. Figure \ref{Y} presents the retrieval performance on each $k$ and $b$ value. Obviously, the larger $k$, the more information contained in $\bar{\mathbb{Y}}$, and the better the final performance. When $k$ is larger than 100, the gain becomes smaller. Therefore, we set $k = 100$ in our experiments, which provides sufficient information.


\begin{table}[t!]
\vspace{-0.15in}
\caption{Ablation study on MS MARCO Dev Doc dataset.}
\vspace{-0.15in}
\label{table2}
\begin{tabular}{ccccc}
\hline
\multirow{2}{*}{Model} & \multicolumn{3}{c}{MARCO Dev Doc} &  \\
 & \multicolumn{1}{c}{MRR@100} & \multicolumn{1}{c}{R@100} & \multicolumn{1}{c}{AQT(ms)}  \\
\hline
IVFFlat & 0.310 & 0.714 & 24 \\
Tree & 0.256 & 0.556 & 5 \\
$+$Joint Optimization & 0.296 & 0.640 & 5 \\
$+$Reorganize clusters & 0.303 & 0.678 & 5 \\
$+$Overlapped clustering & 0.327 & 0.743 & 8\\
\hline
\end{tabular}
\vspace{-5mm}
\end{table}

\subsection{Ablation Study}
In this section, we conduct ablation studies on JTR to explore the importance of different components. We use IVFFlat as the baseline. IVFFlat forms as many clusters as the leaf nodes of the tree. The $nprobe$ of IVFFlat and the beam size of JTR are both set to 10. We use the following four model variants:
\begin{itemize}
		\item[$\bullet$] Tree: The STAR model is used to obtain the document embeddings, and then we build the tree index with MSE loss.
		\item[$\bullet$] $+$Joint Optimization: Joint optimization of query encoders and tree-based indexes using the unified contrastive learning loss function.
		\item[$\bullet$] $+$Reorganize Clusters: Update the clustering of documents, which means the number of overlapped clustering $\lambda=1$.
		\item[$\bullet$] $+$Overlapped Clustering: Documents can be re-occurring in the clustering nodes. For convenience, the number of overlapped clustering $\lambda=2$.
\end{itemize}

Table \ref{table2} shows the $MRR@100$ and $R@100$ on the development set of the MS MARCO Document Ranking task. As the results show, the tree structure significantly reduces the retrieval latency. All of Joint Optimization, Reorganize Clusters and Overlapped Clustering improve the performance of JTR. Tree-based indexes benefit from supervised data directly by Joint Optimization. Reorganize clusters makes the distribution of embeddings more reasonable. Overlapped cluster mines different semantic information in documents. This result demonstrates the effectiveness of our method.

\section{CONCLUSION}
To improve the efficiency of the DR models while ensuring the effectiveness, we propose JTR, which jointly optimizes tree-based index and query encoder in an end-to-end manner. To achieve this goal, we carefully design a unified contrastive learning loss and tree-based negative sampling strategy. Through the above strategies, the constructed index tree possess the maximum heap property which easily supports beam search.
Moreover, for cluster assignment, which is not differentiable w.r.t. contrastive learning loss, we introduce overlapped cluster optimization. We further conducted extensive experiments on several popular retrieval benchmarks. Experimental results show that our approach achieves competitive results compared with widely-adopted baselines. Ablation studies demonstrate the effectiveness of our strategies. Unfortunately, since different indexes have varying degrees of code optimization, we do not report the memory case in our paper. In the future, we will try to jointly optimize PQ and tree-based index to achieve the ``effectiveness-efficiency-memory" tradeoff.

\begin{acks}
This work is supported by the Natural Science Foundation of China (62002194), Tsinghua University Guoqiang Research Institute, Tsinghua-Tencent Tiangong Institute for Intelligent Computing and the Quan Cheng Laboratory.
\end{acks}

\clearpage
\balance
\bibliographystyle{ACM-Reference-Format}
\bibliography{sample-base.bib}
\end{document}